# Effect of Vacancies on the Mechanical Properties of Phosphorene Nanotubes

V. Sorkin* and Y.W. Zhang†

Institute of High Performance Computing, A*STAR, Singapore 138632

## Abstract

Using density functional tight-binding method, we studied the mechanical properties, deformation and failure of armchair (AC) and zigzag (ZZ) phosphorene nanotubes (PNTs) with monovacancies and divacancies subjected to uniaxial tensile strain. We found that divacancies in AC PNTs and monovacancies in ZZ PNTs possess the lowest vacancy formation energy, which decreases with the tube diameter in AC PNTs and increases in ZZ PNTs. The Young's modulus is reduced, while the radial and thickness Poisson's ratios are increased by hosted vacancies. In defective AC PNTs, deformation involves fracture of the intra-pucker bonds and formation of the new inter-pucker bonds at a critical strain, and the most stretched bonds around the vacancy rupture first, triggering a sequence of the structural transformations terminated by the ultimate failure. The critical strain of AC PNTs is reduced significantly by hosted vacancies, whereas their effect on the critical stress is relatively weaker. Defective ZZ PNTs fail in a brittle-like manner once the most stretched bonds around a vacancy rupture, and vacancies are able to significantly reduce the failure strain but only moderately reduce the failure stress of ZZ PNTs. The understandings revealed here on the mechanical properties and the deformation and failure mechanisms of PNTs provide useful guidelines for their design and fabrication as building blocks in nanodevices.

## 1. Introduction

Phosphorene, a two-dimensional (2D) layer of black phosphorus, is a fairly new addition to the rapidly growing family of 2D materials [1–4]. Its semiconducting features [5,6] and unusual thermo-mechanical and transport properties [7] have attracted significant interest in view of its potential applications in flexible electronics [3], thermal management [8], batteries [9], p-n junctions [10], gas sensors [11] and solar-cells [12].

Beside phosphorene, phosphorene-based nanostructures, such as nanoribbons and nanotubes, have also attracted a considerable interest since they can be used as building blocks for nanoelectromechanical systems (nanoresonators, nanoaccelerometers and integrated detection devices), electronic circuits (nanotube-based transistors, field emitters and sensors), smart materials and biomedical applications [5,13], much like graphene nanoribbons and carbon nanotubes. Currently, the study of the electronic, thermal and mechanical properties of phosphorene nanotubes (PNTs) is an active field of research.

---

* Email address: sorkinv@ihpc.a-star.edu.sg
† Email address: zhangyw@ihpc.a-star.edu.sg

Similar to phosphorene, PNTs are direct band gap semiconductors and their electronic properties can be effectively tuned by strain or electric field [14,15]. The puckered structure of phosphorene defines many specific mechanical properties of PNTs, in particular, their strong anisotropy [14,16]. The Young's modulus, radial Poisson's ratio and failure stress of pristine arm-chair (AC) PNTs are considerably larger than those of pristine zig-zag (ZZ) PNTs, whereas the flexural rigidity, thickness Poisson's ratio and failure strain of pristine ZZ PNTs exceed those of pristine AC PNTs [16]. Under applied tensile strain, pristine ZZ PNTs pass through three distinct deformation phases before the ultimate failure, while pristine AC PNTs are monotonously strained to failure induced by the rupture of overstretched bonds [16]. For both pristine AC and ZZ PNTs, the failure strain and failure stress are sensitive, while the Young's modulus, flexural rigidity, and Poisson's ratios are insensitive to the tube diameter [14,16].

Previous studies mainly focused on the mechanical properties of pristine PNTs. However, in real conditions, the presence of defects in PNTs is inevitable. Moreover, various defects can be deliberately introduced into nanotubes during their fabrication to achieve specific functionalities that are absent in pristine nanotubes. Since defects can have profound effects on the electronic, thermal and mechanical properties of PNTs, a closer examination of defects in nanotubes is an essential task from both the practical and fundamental standpoints. The knowledge of the mechanical properties, deformation and failure mechanisms is indispensable since it sets the limits for mechanical strains and stresses in nanoelectromechanical systems that contain defective PNTs, and provides useful guidelines for the design and fabrication of PNTs as building blocks in nanodevices.

In this work, we studied the effects of vacancy defects on the mechanical properties of PNTs: the elastic properties, tensile deformation and failure behavior of PNTs. Two types of vacancies (single and double vacancies) and two primary types of tube geometry (AC and ZZ) were examined. We also investigated how the tube diameter affects the mechanical properties of defective PNTs. In essence, we would like to answer the following closely related questions: What are the structures and formation energies of the hosted vacancies in PNTs? How does the vacancy formation energy depend on the tube radius? In what way are the Young's modulus, flexural rigidity and Poisson's ratio affected by different types of vacancies? What are the failure mechanisms for PNTs with defects? How are the failure strain and failure stress dependent on the type of hosted vacancies? To answer these questions, we carried out density functional tight binding calculations.

## 2. Computational Model

We used the tight-binding (TB) method [17,18] to study the deformation and failure of defective PNTs under uniaxial tensile strain. The TB technique was applied since the density function theory (DFT) is computationally overly expensive for the relatively large defective nanotubes (containing up to ~500 atoms) subjected to large tensile deformations leading to the ultimate failure. Molecular dynamics (MD) simulations cannot be used since a highly-reliable and broadly accepted interatomic potential for phosphorene is unavailable. In addition, MD simulations produce less accurate results since they neglect the electronic degrees of freedom and all associated quantum effects. Hence, the TB technique that yields a suitable balance between MD efficiency and DFT accuracy is the best approach to deal with the above-mentioned questions.

In our simulations of defective PNTs, we applied the density functional tight-binding (DFTB) method [19,20]. The DFTB method was derived from the DFT but used a number of empirical approximations to increase its computational efficiency, while preserving to a great extent the accuracy of the DFT method [20–23]. The substitution of the many-body Hamiltonian of DFT with a parameterized Hamiltonian matrix is the key approximation of the DFTB [20,24]. Quasi-atomic wave functions were constructed by using DFT calculations, and then applied to calculate the Hamiltonian matrix [19]. A few extra terms were added to the Hamiltonian matrix to describe the short-range repulsion, van der Waals and Coulomb interactions [25]. Additionally, the self-consistent charge (SCC) technique was used to substantially improve the description of atom bonding in the DFTB [25]. Recently, the DFTB method were applied to investigate the mechanical properties of phosphorene monolayer, nanoribbons and pristine nanotubes, as well as grain boundaries in phosphorene [14,16,26–33]. In our simulations, we used the open-source quantum mechanical simulation package "DFTB+" [25,34].

As a starting point, we optimized the unit cell of phosphorene obtained by the DFT method [12]. Then a phosphorene sheet was constructed and a vacancy was introduced by removing a single atom or a pair of adjacent atoms. The initial atomistic configuration around the created vacancy was adjusted by using the results of DFT calculations [35]. Subsequently, the constructed phosphorene sheet with a vacancy was rolled along the AC or ZZ direction into an AC PNT or a ZZ PNT (see Figure 1). The atomistic configuration around the lattice site with a missing single atom in AC PNTs (obtained after relaxation of the tube geometry) is highlighted in Figure 1(a). It is seen that the two adjacent pentagon-nonagon atomic rings sharing the common bond constitute the monovacancy, and the three neighboring pentagon-octagon-pentagon atomic rings sharing the two mutual bonds comprise the divacancy. There are two distinct types of divacancies (DV(585)A and DV(585)B) as illustrated in Figure 1(b, c), respectively. The corresponding monovacancy and two variants of divacancies in ZZ PNTs have the similar geometrical structure (see Figure 1(f-i)), although their orientations with respect to the tube axis are different. The shell bonds connecting phosphorous atoms (P-atoms) within the inner and outer tube shells, as well as the bridge bonds connecting the inner and outer shell atoms in AC and ZZ PNTs are shown in Figure 1(e, j), correspondingly.

PNTs can be specified either by the number of phosphorene unit cells along the tube circumference or by the tube diameter. By changing the number of unit cells along the tube circumference from N=12 to N=20, we constructed AC PNTs with the tube diameter ranging from D=18.9 Å to D=30.1 Å, and ZZ PNTs with the diameter varying from D=15.4 Å to D=24.3 Å. The length of the constructed AC and ZZ PNTs was chosen to be L=22.5 Å and L=28.4 Å, respectively. The total number of atoms in the PNTs is varied in our simulations from ~300 to ~550.

We studied the mechanical response of defective PNTs subjected to uniform uniaxial tensile strain applied alongside the tube axis (oriented along the X direction as shown in Figure 1). Since we applied periodic boundary conditions in all the directions, a vacuum slab with the thickness of $\Delta w$=20 Å was added along the Y and Z directions to prevent the self-interaction of PNTs (due to the periodic boundary conditions along the Y and Z directions). The k-point set for the Brillouin-zone integration was selected using the Monkhorst-Pack method [36]. The Monkhorst-Pack grid [36] with an 8x2x2 sampling set was used for Brillouin-zone integration. Following the DFTB studies for pristine PNTs [14,16], the s- and p-orbitals were specified for each P-atom. The Slater-Koster files [17] for P-atoms were selected from the 'MATSCI' set [37,38].

Uniaxial tensile strain was applied to the PNTs quasi-statically along the tube axis (X-axis) at zero temperature. The tensile strain was gradually incremented by a small step of $\delta\varepsilon$=0.01. Subsequently, the geometry of the defective PNTs was relaxed by minimizing the total energy at a given tensile strain with the conjugate gradient method [39]. The SCC calculations were carried out at each step of the energy minimization. We estimated the nominal stress, as explained in [12], by calculating the force per unit of the effective cross-sectional area, which was defined as the difference between the areas of the outer and the inner circles outlined by the outer and inner tube shells [16]. The volume of the nanotube was calculated as a product of its length and its effective area.

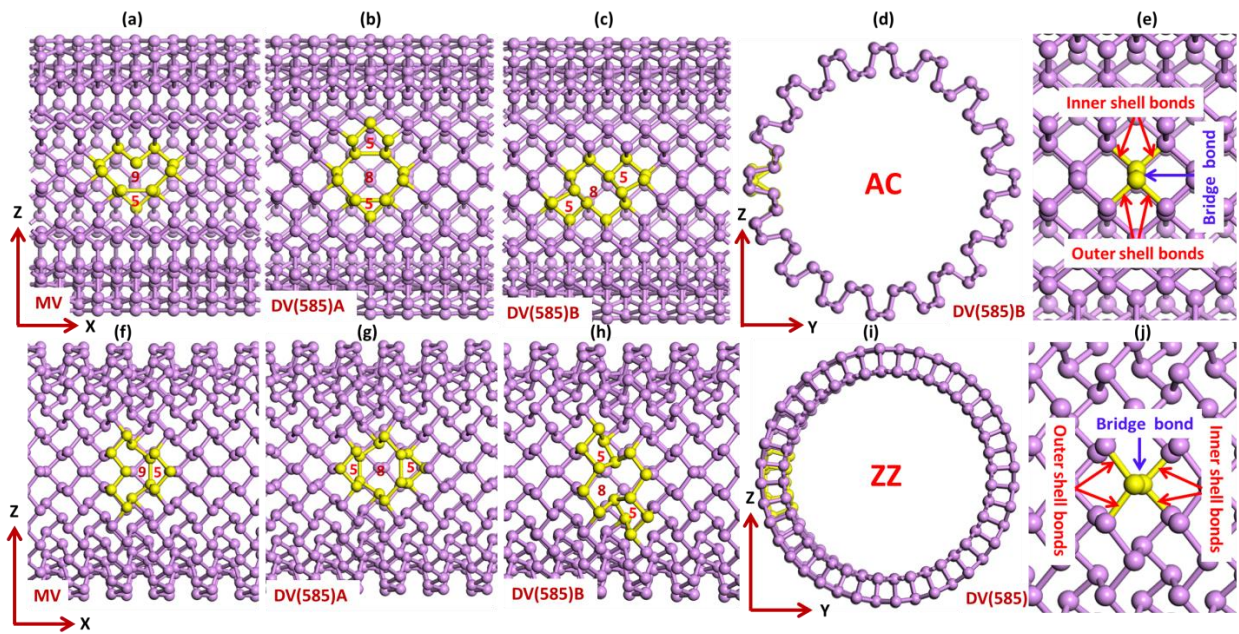

Figure 1: (a-d) Side view of an AC PNT oriented along the X-axis with a MV(59) monovacancy (a), a DV(585)A divacancy (b) and a DV(585)B divacancy (c). The diameter of the AC PNT is D = 24.5Å. Atoms around the vacancy sites are highlighted and the atomic rings (pentagon, octagon and nonagon) are indicated. (e) Orientation of the bridge, inner and outer shell bonds in the AC PNT. (f-i) Side view of a ZZ PNT oriented along the X-axis with a MV(59) monovacancy (f), a DV(585)A divacancy (g) and a DV(585)B divacancy (h). The diameter of the ZZ PNT is D = 19.8Å. Atoms around the vacancy sites are highlighted and the atomic rings (pentagon, octagon and nonagon) are indicated. (j) Orientation of the bridge, inner and outer shell bonds in the ZZ PNT.

## 3. Results and Discussion

### 3.1 Defect structure and energetics

First, we optimized the geometry of the constructed PNTs with defects. The optimized geometries of AC and ZZ PNTs containing single and double vacancies are shown in Figure 1. The atomistic configuration around a monovacancy in AC PNTs is presented in Figure 1(a), where the adjacent pentagon-nonagon rings are highlighted. The two variants of divacancies: DV(585)A and DV(585)B are shown in Figure 1(b, c), respectively. The octagon and two pentagon rings are highlighted. In the DV(585)A divacancy, the most strained bonds are shared by the octagon and pentagon rings. The corresponding bonds of the DV(585)B divacancy are slightly skewed with respect to the tube axis (see Figure 1(c)). Vacancies in ZZ PNTs have the similar geometrical structure; however, their orientations with respect to the tube axis are different. As can be seen in Figure 1(h-i), the most strained bonds shared by the pentagon-octagon rings of the DV(585)A divacancy are perpendicular to the tube axis.

The spatial energy distribution (energy per atom) around the hosted vacancies in the reference configuration (at zero strain: ε=0) is shown in Figure 2. The energy of atoms around the vacancies is evidently higher. Around the AC PNT monovacancy, the pair of atoms shared by the pentagon and nonagon rings and the atom with a dangling bond attains the highest energy (see the red atoms in Figure 2(a)). Around the DV(585)A divacancy, the two inner shell atoms bound by a bond and shared by the pentagon and octagon rings have the highest energy (see the red atoms in Figure 2(b)). Equally, around the DV(585)B divacancy, the highest energy is attained by the pair of atoms shared by the neighboring pentagon and octagon rings (see Figure 2(c)) although their energy is smaller than those of other types of vacancies. A similar energy distribution is also found around the vacancies in ZZ PNTs (see Figure 2(d-f)): around the monovacancy, the highest energy is attained by the atom with a dangling bond, and around the divacancy, the highest energy is attained by the two nearby outer shell atoms shared by the pentagon-octagon rings. As can be seen in Figure 2(d-f), the energy of the inner shell atoms in ZZ PNTs is lower than that of the outer shell atoms. The energy difference is associated with the particular geometry of ZZ PNTs, where the phosphorene puckers are oriented along the tube circumference [16]. The smaller the tube diameter is, the larger the energy difference is.

Next, we calculated the vacancy formation energy according to:

$$E_{vac} = E_{tot} - \frac{N}{N_p} E_{p,tot}$$

where $E_{tot}$ is the potential energy of a defective PNT containing *N* atoms, while $E_{p,tot}$ is the potential energy of the corresponding defect-free PNT with the same radius containing $N_p$ atoms ($N = N_p - n$, where *n* is the number of removed atoms: *n*=1 for monovacancy and *n*=2 for divacancy).

The vacancy formation energy as a function of the tube diameter is shown in Figure 3(a) for AC PNTs and Figure 4(a) for ZZ PNTs. In AC PNTs, the DV(585)B divacancy possesses the lowest vacancy formation energy, while the DV(585)A divacancy possesses the highest one at the small tube diameters. At the large tube diameters, the vacancy formation energy of DV(585)B and monovacancy in AC PNTs become similar (see Figure 3(a)). At large diameters, the vacancy formation energies of DV(585)B and monovacancy in AC PNTs become comparable (see Figure 3(a)). For ZZ PNTs, the vacancy formation energy of the monovacancy is the lowest one, while that of the DV(585)A divacancy is the highest one, regardless of the tube diameter (see Figure 4(a)). In AC PNTs, the vacancy formation energy decreases with increasing the tube diameter: the larger the tube diameter is, the smaller the vacancy formation energy is. In contrast to AC PNTs, the vacancy formation energy in ZZ PNTs increases with increasing the tube diameter.

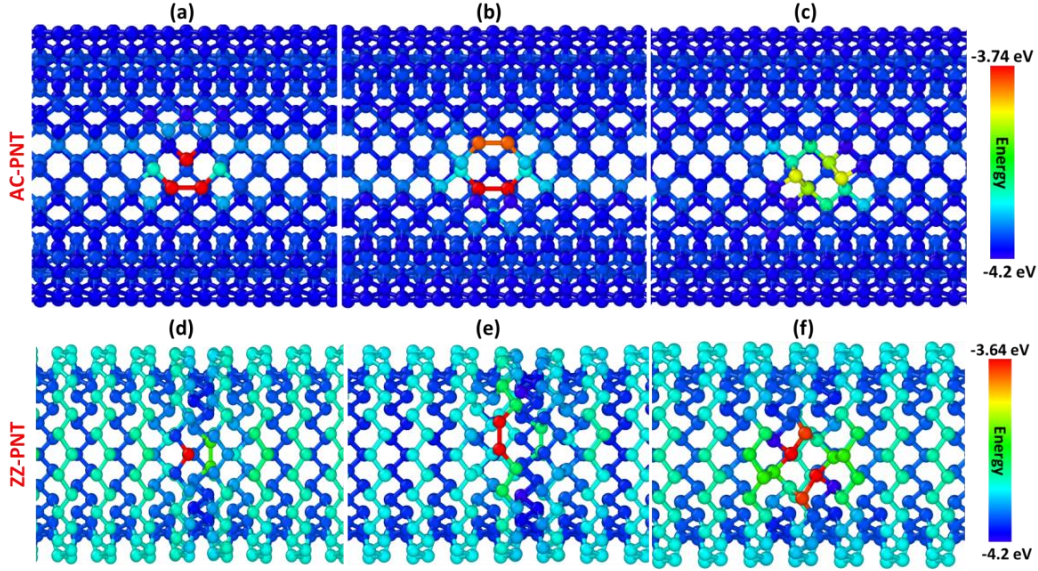

Figure 2: The spatial energy distribution in AC (upper panel) and ZZ (lower panel) PNTs with the monovacancies (a, d) and two variants of divacancies: DV(585)A (b, e) and DV(585)B (c, f). Color specifies energy per atom. Energy range is indicated by color bars on the right. The diameter of the AC PNT is D = 24.5 Å and that of the ZZ PNT is D = 22.1 Å.

To understand why the vacancy formation energy decreases with the increase of the tube diameter in AC PNTs but increases in ZZ PNTs, we examined the energy distribution around a DV(585)A divacancy in narrow and wide PNTs. For the clarity of presentation, instead of using the energy per atom, we plot the energy difference per atom by subtracting the energy of the corresponding inner and outer shell atoms in a pristine nanotube with the same diameter. This energy difference represents an individual atomic contribution to the vacancy formation energy. The energy difference around the DV(585)A divacancy in AC PNTs is shown in Figure 3(c-f)  (the numerical values are indicated; for the symmetry reasons, the energy contributions are specified only for half of the highlighted atoms). The energy difference is shown separately for the inner and outer tube shells. Whereas the energy difference for the inner shell atoms is visualized, the color of the outer shell atoms remained fixed (and vice versa). For comparison, the energy difference around the DV(585)A divacancy in phosphorene monolayer is also shown in Figure 3(b).

For both the narrowest (D=18.9Å) and widest (D=30.1Å) AC PNTs, the main contribution to the vacancy formation energy comes from the octagon atoms (in particular from the pairs of atoms bound by the shared pentagon-octagon bonds, see Figure 3(c-f)). The energy contributions of the inner and outer shell atoms located around the DV(585)A divacancy are different. In the narrow AC PNT, the contribution of the inner shell atoms ($\Delta E_{in}$ = 1.18 eV) is larger than the contribution of the outer shell atoms ($\Delta E_{out}$ = 1.16 eV). The corresponding total vacancy formation energy $\Delta E_v$ = 2.34 eV is larger than that of the monolayer $\Delta E_v$ = 2.24eV. In contrast, in the wide AC PNT, the contribution of the inner shell atoms ($\Delta E_{in}$ = 1.12 eV) is smaller than the contribution of the outer shell atoms ($\Delta E_{out}$ = 1.15 eV). The total vacancy formation energy $\Delta E_v$ = 2.27 eV is lower than that of the narrow AC PNT. Evidently, this is an effect of the tube curvature: the bonds oriented along the tube circumference are more stretched in the narrow nanotubes than in the wide ones [16], particularly around a vacancy. Hence, the contribution of the atoms linked by these bonds to the vacancy formation energy is larger. As a result, the vacancy formation energy in AC PNTs decreases with the increase of the tube diameter.

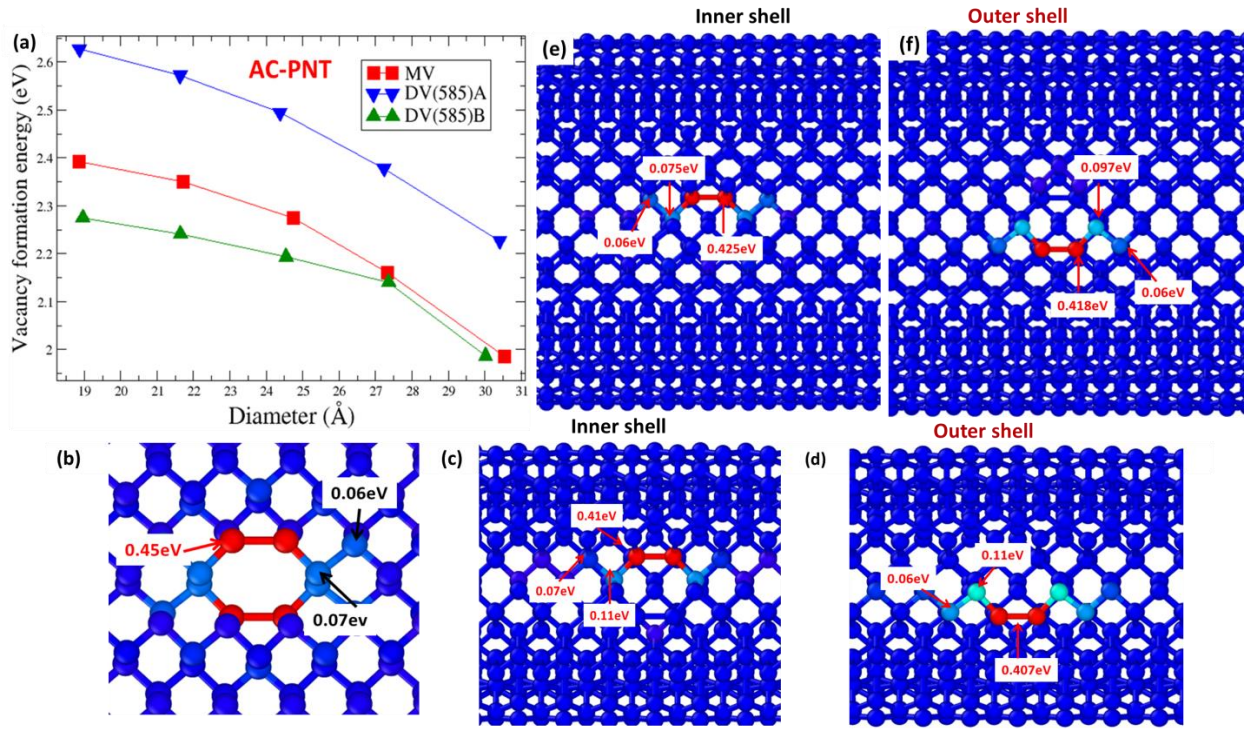

Figure 3: (a) The formation energy of a monovacancy (red squares), a DV(585)A (blue triangles down) and a DV(585)B (green triangles up) divacancy in AC PNTs as a function of tube diameter. (b) The energy difference per atom around the DV(585)A divacancy in a phosphorene monolayer. (c, d) The energy difference per atom around the DV(585)A divacancy in the inner (c) and outer (d) shell of a narrow AC PNT (D=18.9Å). (e, f) The energy difference per atom around the DV(585)A divacancy in the inner (e) and outer (f) shell of a wide AC PNT (D=30.1Å). The numerical values for the energy difference specified in (b-f) are obtained by subtracting the energy per atom in a defect-free monolayer or a nanotube with the same diameter.

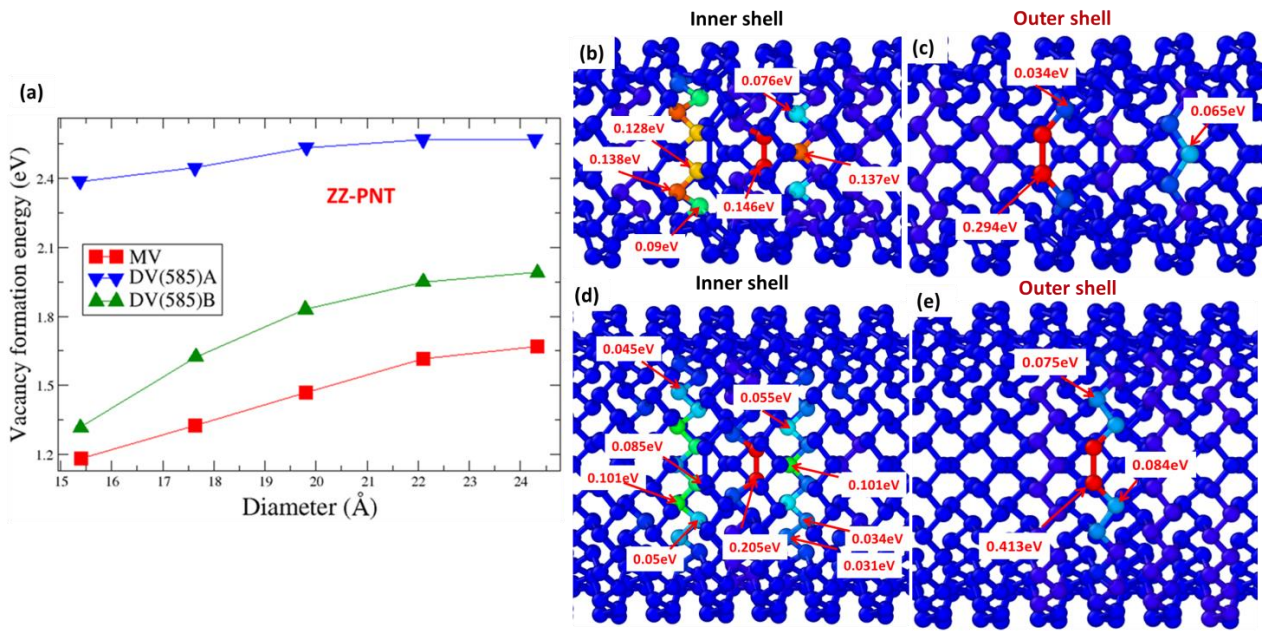

Figure 4: (a) The formation energy of a monovacancy (red squares), a DV(585)A (blue triangles down) and a DV(585)B (green triangles up) divacancy in ZZ PNTs as a function of tube diameter. (b,c) The energy difference per atom around the DV(585)A divacancy in the inner (b) and outer (c) shell of a narrow ZZ PNT (D=15.4Å). (d,e) The energy difference per atom around the DV(585)A divacancy in the inner (d) and outer (e) shell of a wide ZZ PNT (D=24.3Å). The numerical values for the energy difference specified in (b-e) are obtained by subtracting the energy per atom in a defect-free monolayer or a nanotube with the same diameter.

In the same way, the individual atomic contributions to the vacancy formation energy in the narrowest (D=15.4Å) and the widest (D=24.3Å) ZZ PNTs were examined (see Figure 4(b-e)). As can be seen in Figure 4(b-e), the contributions of the inner and outer shell atoms are different. In contrast to AC PNTs, the number of the inner shell atoms in ZZ PNTs contributing to the vacancy formation energy is significantly larger (compare Figure 3(c, e) and Figure 4 (b, d)). In the narrow ZZ PNT, the total vacancy formation energy is $\Delta E_v$ = 2.01 eV, with $N_i$=11 inner shell atoms (compare to $N_i$=6 atoms in the narrow AC PNT) contributing $\Delta E_{in}$ = 1.29 eV, while $N_0$=5 outer shell atoms contributing $\Delta E_{out}$ = 0.72 eV to the vacancy formation energy (see Figure 4 (b, c)). In the wide ZZ PNT, the total vacancy formation energy is $\Delta E_v$ = 2.75 eV, with $N_i$=17 inner shell atoms contributing $\Delta E_{in}$ = 1.31 eV, while $N_0$=5 outer shell atoms contributing $\Delta E_{out}$ = 1.44 eV to the vacancy formation energy (see Figure 4 (d, e)). Evidently, the main energy contribution comes from the outer shell atoms bound by the pentagon-octagon bond. Their energy contribution is almost twice as large as that of the inner shell atoms sharing the same bond since the ZZ tube curvature can be accommodated only when the outer shell bonds are more strained than the inner shell ones. The outer shell pentagon-octagon bond of the narrow tube is stretched by $\Delta\varepsilon_s \approx 8\%$, while the same inner shell bond only by $\Delta\varepsilon_s \approx 5\%$. The outer shell pentagon-octagon bond of the wide nanotube is stretched by $\Delta\varepsilon_s \approx 12\%$, while the same inner shell bond only by $\Delta\varepsilon_s \approx 8\%$. Hence, the contribution associated with the strained pentagon-octagon bonds increases with the tube diameter. The larger the diameter of ZZ PNTs is, the larger the contribution of the inner and outer shell atoms to the vacancy formation energy is. Therefore, the vacancy formation energy in ZZ PNTs increases with the tube diameter.

## 3.2 Flexural rigidity, Young's modulus and Poisson's ratios

We also examined the effect of vacancies on the mechanical properties of defective PNTs: flexural rigidity, Young's modulus and Poisson's ratios. First, the effect of vacancies on the flexural rigidity of PNTs was investigated. According to the linear elasticity theory [40], the strain energy ($E_s$), associated with the curved tube geometry, defined as a difference between the energy per atom of the tubular and planar (monolayer) structure, decreases with increasing the tube diameter (D) as: $E_s = \frac{\delta}{D^2}$, where $\delta$ is the flexural rigidity. The strain energy was calculated according to: $E_s = \frac{E_{PNT}}{N} - E_{at}$, where $E_{PNT}$ is the total potential energy of the nanotube, N is the number of the nanotube atoms and $E_{at}$ is the energy per atom in phosphorene sheet (monolayer).

We found that much like in defect-free PNTs [16], the strain energy associated with tube rolling is lower for the defective AC PNTs than for the defective ZZ PNTs: the difference is especially apparent for the PNTs with small diameters. The lower strain energy of AC PNTs implies that they are more energetically favorable than ZZ PNTs [14,15]. The values of the flexural rigidity obtained by fitting the strain energy of the defect-free and defective nanotubes are listed in Table 1. The effect of both single and double vacancies on the flexural rigidity is weak: the hosted vacancies (mostly divacancies) slightly increase the flexural rigidity. The tube structure is locally altered around the introduced vacancies; therefore the larger strain energy is associated with bond stretching around the defects, which leads to the larger flexural rigidity.

The Young's modulus (*Y*) of defective PNTs was calculated as a second derivative of the total potential energy with respect to strain (taken in the reference configuration at ε=0):

$$Y = \frac{1}{V_0}\left(\frac{\partial^2 E_{PNT}}{\partial \varepsilon^2}\right)_{\varepsilon=0}$$

Here $V_0$ is the tube volume measured at zero strain. The second derivative of $E_{PNT}$ was calculated by subjecting PNTs to infinitesimally small compressive and tensile strains (-3% ≤ ε ≤ 3%), with subsequent optimization of the tube geometry at a given strain. The obtained $E_{PNT}$ as a function of uniaxial strain was fitted to a second order polynomial, which was used to calculate the second derivative.

The calculated Young's modulus of pristine and defective AC and ZZ PNTs as a function of the tube diameter is shown in Figure 5(a, b), respectively. As can be seen in Figure 5(a), the Young's modulus of AC PNTs is reduced by the hosted vacancies. A DV(585)A divacancy has the strongest effect on the Young's modulus ($\Delta Y/Y_0 \approx$ -8%), while a monovacancy has the weakest one ($\Delta Y/Y_0 \approx$ -3%). The Young's modulus of pristine and defective AC PNTs gradually decreases with the increase in the tube diameter.

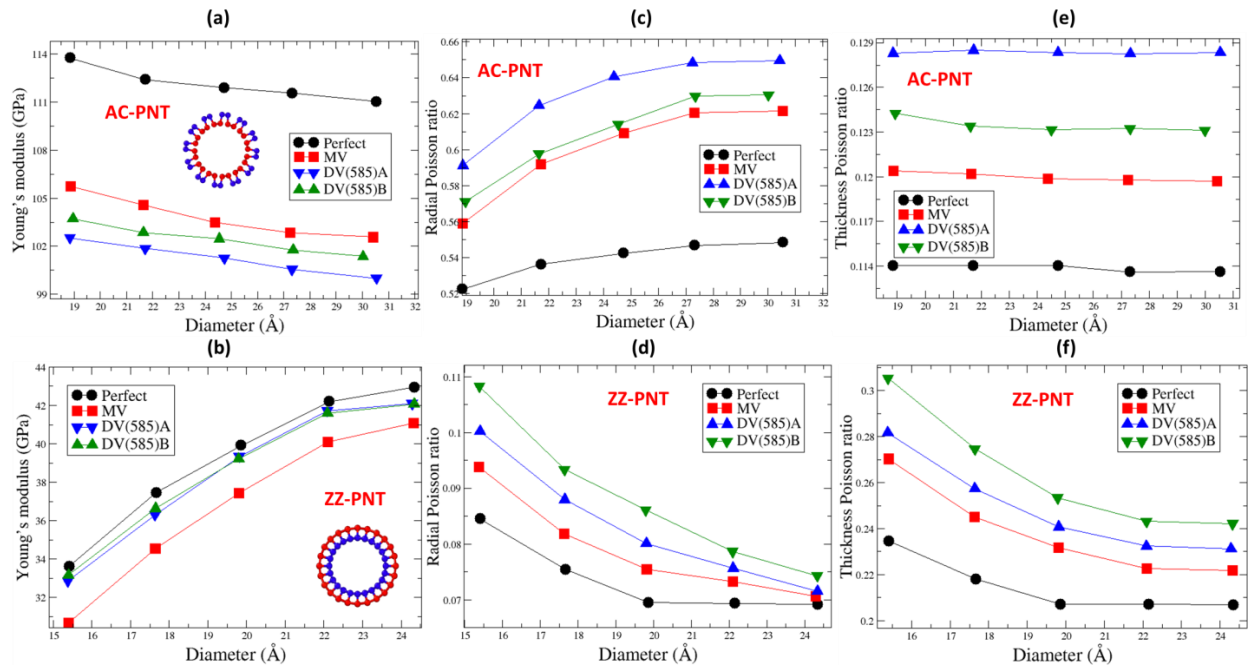

Figure 5: (a, b) The Young's modulus of AC (a) and ZZ (b) PNTs with a monovacancy (red squares), a DV(585)A (blue triangles down) and a DV(585)B (green triangles up) divacancy as a function of tube diameter. The Young's modulus of defect-free PNTs is specified by the black circles. (c, d) The radial Poisson's ratio as a function of tube diameter for AC (c) and ZZ (d) PNTs with a monovacancy (red squares), a DV(585)A (blue triangles up) and a DV(585)B (green triangles down) divacancy, as well as without vacancies (black circles). (e, f) The thickness Poisson's ratio as a function of tube diameter for AC (e) and ZZ (f) PNTs with a monovacancy (red squares), a DV(585)A (blue triangles up) and a DV(585)B divacancy (green triangles down), as well as without vacancies (black circles).

The Young's modulus of ZZ PNTs is also reduced by the hosted vacancies (see Figure 5(b)), but in contrast to AC PNTs, it is more affected by a monovacancy ($\Delta Y/Y_0 \approx$ -7%) rather than by a divacancy ($\Delta Y/Y_0 \approx$ -3%). Furthermore, the Young's modulus of ZZ PNTs enlarges with the increasing tube diameter (when the tube diameter is rather small: D ≲ 20Å) and slowly converges to a constant (when the tube diameter is sufficiently large; see Figure 5(b)).

We also calculated the radial Poisson's ratio ($\nu_D$), which is defined as the ratio of the transverse contraction strain $\frac{\Delta D}{D}$ to the longitudinal extension strain $\frac{\Delta L}{L}$ of PNTs, and can be obtained from:

$$\frac{\Delta D}{D} = -\nu_D \frac{\Delta L}{L},$$

where $D$ is the tube diameter and $L$ is the tube length. The Poisson's ratio for the tube thickness ($\nu_t$) was obtained in the same way, by replacing the diameter $D$ with the tube thickness ($t$), which was calculated as the difference between the outer and inner tube radii. The Poisson's ratios $\nu_D$ and $\nu_t$ are both positive since the tube elongation along its axis is accompanied by reduction in its diameter and thickness (see Figure 5(c-f)).

In Figure 5(c), we plot $\nu_D$ as a function of tube diameter for defective AC PNTs. The hosted vacancies increase the radial Poisson's ratio: $\nu_D$ is mainly affected by DV(585)A divacancies, while the effect of DV(585)B divacancies and monovacancies is weaker. The radial Poisson's ratio also increases with tube diameter: $\nu_D$ of the narrowest nanotube (D=18.9Å) with a DV(585)A divacancy increases by $\Delta \nu_D / \nu_{D_0} \approx 13\%$, while $\nu_D$ of the widest nanotube (D=30.1Å) increases by more than $\Delta \nu_D / \nu_{D_0} \approx 20\%$ (see Figure 5(c)). The effect of vacancies on the radial Poisson's ratio of ZZ-PNTs is similar (see Figure 5(d)): $\nu_D$ is mostly affected by DV(585)B divacancies, while the effect of DV(585)A divacancies and monovacancies is less significant. The impact of the hosted vacancies on $\nu_D$ lessens with tube diameter: $\nu_D$ of a ZZ nanotube with the smallest diameter (D=15.4Å) increases by $\Delta \nu_D / \nu_{D_0} \approx 26\%$ due to the DV(585)B divacancy, while $\nu_D$ of a ZZ nanotube with the largest diameter (D=24.3Å) increases only by $\Delta \nu_D / \nu_{D_0} \approx 5\%$ (see Figure 5(d)).

Vacancies also increase the thickness Poisson's ratio of defective PNTs. The thickness Poisson's ratio of AC PNTs is mostly affected by DV(585)A divacancies, while $\nu_t$ of ZZ PNTs by DV(585)B divacancies. For both pristine and defective AC PNTs, $\nu_t$ is almost independent of tube diameter (see Figure 5(e)): in AC PNTs with a DV(585)A divacancy, $\nu_t$ increases by $\Delta \nu_t / \nu_{t_0} \approx 12\%$ for the smallest tube diameter and by $\Delta \nu_t / \nu_{t_0} \approx 10\%$ for the largest one. In contrast, $\nu_t$ of ZZ PNTs decreases significantly with the tube diameter (see Figure 5(f)): in ZZ PNTs with a DV(585)B divacancy, $\nu_t$ increases by $\Delta \nu_t / \nu_{t_0} \approx 24\%$ for the smallest diameter and only by $\Delta \nu_t / \nu_{t_0} \approx 8\%$ for the largest one.

We listed the values of the Young's modulus, flexural rigidity and Poisson's ratios calculated for the pristine and defective PNTs in Table 1. For the Young's modulus, flexural rigidity and Poisson's ratios, which moderately depend on the tube diameter, we used the converged values extrapolated for the large tube diameters.

Table 1: The Young's modulus, flexural rigidity and Poisson's ratios of the pristine and defective PNTs

| Defect type | Young's modulus (GPa) | | Flexural rigidity (eV·nm$^2$/atom) | | Radial Poisson's ratio | | Thickness Poisson's ratio | |
|---|---|---|---|---|---|---|---|---|
| | AC | ZZ | AC | ZZ | AC | ZZ | AC | ZZ |
| Pristine | 111.1 | 42.9 | 0.0192 | 0.0710 | 0.47 | 0.07 | 0.11 | 0.21 |
| Monovacancy | 102.5 | 41.09 | 0.0237 | 0.0713 | 0.62 | 0.071 | 0.12 | 0.22 |
| Divacancy (A) | 100.1 | 42.1 | 0.0213 | 0.0724 | 0.65 | 0.072 | 0.13 | 0.23 |
| Divacancy (B) | 101.4 | 42.1 | 0.0224 | 0.0711 | 0.63 | 0.074 | 0.12 | 0.24 |

## 3.3 Tensile deformation and failure pattern

### Deformation and failure of AC PNTs

Next, we investigated the tensile deformation and failure of AC PNTs with single and double vacancies (see Figure 6). The defective AC PNTs were gradually stretched along the axial direction. The strain energy as a function of applied tensile strain for an AC PNT with a DV(585)B divacancy is shown in Figure 6(a). The strain energy increases initially with applied strain, but above a critical tensile strain, a sequence of the abrupt drops in the strain energy are observed. Each abrupt drop indicates a partial release of the strain energy associated with bond breaking and bond remaking between nearby puckers. The last drop in the strain energy is due to the ultimate fracture of the defective AC PNT. In what follows, we describe in detail the structural transformations accompanying the each strain energy drop (as indicated by the inserted numbers in Figure 6(a)).

When AC PNTs are stretched along its axis, the distance between the neighboring puckers along the circumferential direction decreases since the axial elongation of the tube is accompanied by its transverse contraction. In Figure 6(b), we plot the distance between a pair of adjacent puckers in pristine and defective AC PNTs (of the same tube diameter) as a function of uniaxial tensile strain. For the sake of clarity, we present here the results for AC PNTs with a DV(585)B divacancy; the similar results were also obtained for monovacancies and DV(585)A divacancies.

In a defective AC PNT, a pair of atoms was selected in the vicinity of a divacancy (see A and B atoms in the inset of Figure 6(b)). A similar pair of atoms was arbitrarily selected in the inner shell of a pristine AC PNT with the same diameter (see the pair of P-atoms in the inset of Figure 6(b)). As can be seen in Figure 6(b), both the |AB| and |PP| distances decrease almost linearly with applied tensile strain. Since the initial |AB| distance is smaller, it decreases faster and reaches the bond length limit ($l_b \approx 3$Å [16]) first. At $\varepsilon_{cr} \approx 0.16$, a new bond is formed between the pair of A and B atoms, while at the same strain the |PP| distance is still larger than the bond length limit (see Figure 6(b)). The pristine AC PNT can be stretched almost twice as long as the defective one before a new bond can be formed between the pair of P-atoms (see the sharp drop in the |PP| distance at $\varepsilon_{cr} \approx 0.28$ in Figure 6(b)).

Bond formation between A and B atoms initiates a sequence of bond breaking within the puckers and bond remaking between the neighboring puckers, which results in the partial release of the accumulated strain energy (see the ❶→❷ transition in Figure 6(a)). Since the distance between the inner shell atoms decreases slightly faster than the distance between the outer shell atoms, the new inter-pucker bonds are mostly formed between the inner shell atoms. The fraction of the new bonds in the inner shell is f≈5.7% at $\varepsilon_{cr}=0.16$, while in the outer shell it is only f≈2.5% (see Figure 6(c)). The total fraction of the new bonds is plotted in Figure 6(d) as a function of tube diameter. It is higher in the nanotubes with small diameters, because the smaller the tube diameter is, the closer the adjacent puckers are and the higher the likelihood is to form a new bond. As the tube diameter increases, the total fraction becomes a constant independent of the tube diameter. The total fraction depends also on the type of vacancy hosted by AC PNTs (see Figure 6(d)). The presence of dangling bonds leads to the highest fraction of the new inter-pucker bonds around a monovacancy. The fraction of the new bonds formed in the inner shell is ~3.5 larger than that of the outer shell. The differences in the geometrical structure of two types of divacancies have a negligible effect on the fraction of the new inter-pucker bonds around them (see Figure 6(d)). The total fraction of the new inter-pucker bonds in pristine AC PNTs is considerably larger than in defective ones. At the critical strain, the new bonds in a pristine AC PNT are distributed uniformly all over the tube [16], while in a defective nanotube, they are mainly located around a vacancy (see Figure 7(c)).

The two atomistic snapshots taken before and after the first drop in the strain energy are presented in Figure 6(e, f). The new inter-pucker bonds are located along an arc starting at the divacancy and crossing the inner tube shell diagonally. A close inspection of the defective nanotube reveals a sequence of the alternating hexagons and decagons being formed along the tube diagonal. The pattern is similar to that appearing in phosphorene nanoribbons with zigzag edges at a critical tensile strain, where the new inter-pucker bonds are formed along the specific diagonal directions [41].

The atomistic configuration around the hosted divacancy changes at the critical strain. As shown in Figure 6(f), the number of atoms in close vicinity of the divacancy site increases from $N_v$=8 to $N_v$=16. At the critical strain, the divacancy octagon merges with the two nearby hexagons when their commonly shared bonds concurrently rupture. Although the energy of the extended divacancy atoms increases, the effect is negligible for the entire system.

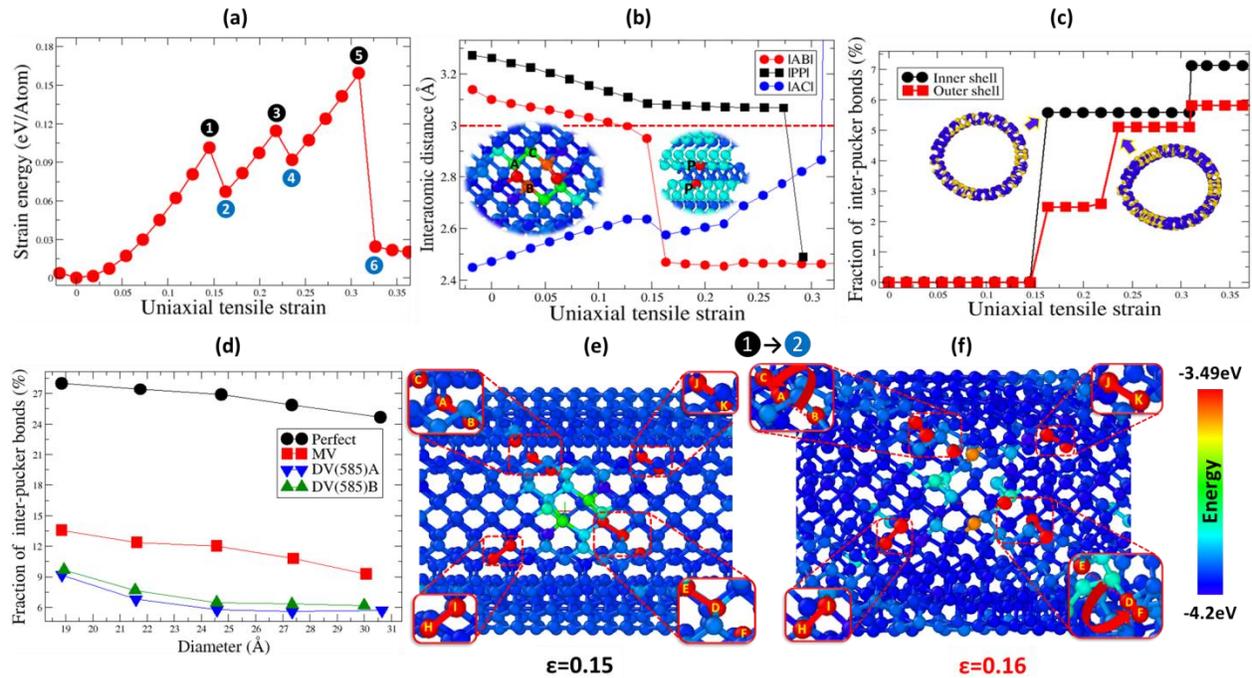

Figure 6: The deformation and failure of defective AC PNTs: (a) The strain energy of an AC PNT (D = 24.5 Å) with a DV(585)B divacancy as a function of applied tensile strain. The inserted numbers indicate the variations in the strain energy associated with the structural transformations. (b) The distance between a pair of next nearest neighbors located in adjacent puckers as a function of applied strain. The |AB| (red circles) is the distance between the pair of A and B atoms near the vacancy, while the |PP|(black squares) is the distance between the arbitrarily chosen pair of P-atoms in adjacent puckers of a pristine AC PNT with the same diameter. For comparison, the bond length between the two nearest neighboring A and C atoms located

around the divacancy |AC| (blue circles) is shown. The red dotted line indicates the maximal bond length. (c) The fraction of the new inter-pucker bonds in the inner (black squares) and the outer (red circles) tube shells as a function of applied tensile strain. Atoms connected by the inter-pucker bonds are highlighted in the inset plots. (d) The total fraction of the new inter-pucker bonds in defect-free (black circles) and defective AC PNTs with monovacancies (red squares), DV(585)A (blue triangles down) and DV(585)B (green triangles up) divacancies as a function of tube diameter. (e, f) Two successive atomistic configurations taken before (e) and after (f) the first drop (❶→❷) in the strain energy at ε=0.15 and $ε_{cr}$=0.16, respectively. Color indicates energy per atom. Energy range is specified by color bar on the right. The highlighted triplets of atoms {A, B, C} and {D, E, F} (shown in the zoomed-up inset plots) are involved in bond breaking-and-remaking process, while the bonds connecting the {H, I} and {J, K} atoms remain intact.

Color mapping allows one to visualize the spatial energy distribution around the divacancy before and after the first structural transformation in the defective AC PNT. The atoms forming the new inter-pucker bonds have a slightly higher energy than the nearby atoms with the original bond connectivity. Even so, the formation of the new bonds and relaxation of the original ones lead to a substantial drop in the strain energy. To make evident that the strain energy decreases after the transformation, we arbitrarily selected a few inner shell atoms forming the new inter-pucker bonds (see A, B, C, D, E and F atoms in the zoomed-up inset plots in Figure 6(e, f)). Before the transition (just below the critical strain) the A-B and E-D atoms are bound: the length of their bonds is $l_{|AB|}$=2.86Å and $l_{|ED|}$=2.82Å, respectively. These bonds are highly strained since their length exceeds significantly the equilibrium bond length ($l_b$=2.48Å [16]) of the inner shell bonds at zero strain in a pristine AC PNT with the same diameter. Just above the critical strain, these bonds are broken and the new bonds (|AC| and |DF|) are formed (see Figure 6(f)). The length of the new inter-pucker bonds is $l_{|AC|}$=2.45Å and $l_{|DF|}$=2.47Å, respectively. Their lengths are much closer to the equilibrium bond length; consequently the strain energy associated with the new bonds is lower than that of the original ones. The bonds preserving the initial bond connectivity relax partially at the critical strain. The two randomly selected bonds between H-I and J-K atoms preserving the original bond connectivity are highlighted in Figure 6(e, f). The length of the bonds just below and above the critical strain is $l_{|HI|}$: 2.64Å→2.57Å and $l_{|JK|}$: 2.7Å→2.6Å, respectively. Evidently, above the critical strain, the length of the selected bonds is closer to the equilibrium bond length; as a result, the strain energy associated with these bonds is reduced. The abrupt drop in the strain energy is a collective effect of a number of bonds involved in the processes of bond breaking-and-remaking and bond relaxation at the critical strain.

The local modification of bond connectivity at the critical strain is an efficient way to accommodate the imposed tensile strain and to reduce the strain energy. In an AC PNT unit cell (see Figure 1(e)) subjected to uniaxial tensile strain, the two outer and two inner shell bonds (oriented along the direction of applied strain) are stretched, while the bridge bond (oriented perpendicularly to the direction of applied strain) is compressed. In a ZZ PNT unit cell (see Figure 1(j)), the bridge bond (oriented along the direction of applied strain) is the most stretched bond [16], whereas the shell bonds (oriented at an angle to the direction of applied strain) are less strained. Stretching of the four shell bonds in the AC PNT unit cell is energetically more expensive than stretching of the single bridge bond in the ZZ PNT unit cell, which can also partially rotate as a hinge to reduce the energy cost of tensile deformation. Therefore, local bond reorientation is an energetically favorable process, which can accommodate the imposed tensile strain at a lower energy cost. However, bond reorientation can happen only when adjacent puckers in AC PNTs are sufficiently close to make bond formation possible.

After the first structural transformation, the AC PNT can be stretched up to ε=0.25, at which the strain energy drops again (see the ❸→❹ transition in Figure 6(a)). At this strain, the new inter-pucker bonds are predominantly formed between atoms of the outer shell since the distance between them becomes sufficiently short (see Figure 6(c)). The new bonds are formed in the same region as the previously formed bonds. The second transition completes the first one: the new bonds transform the previously formed sequence of the alternating hexagons and decagons into the sequence of hexagons containing only the atoms with modified bond connectivity (see Figure 7).

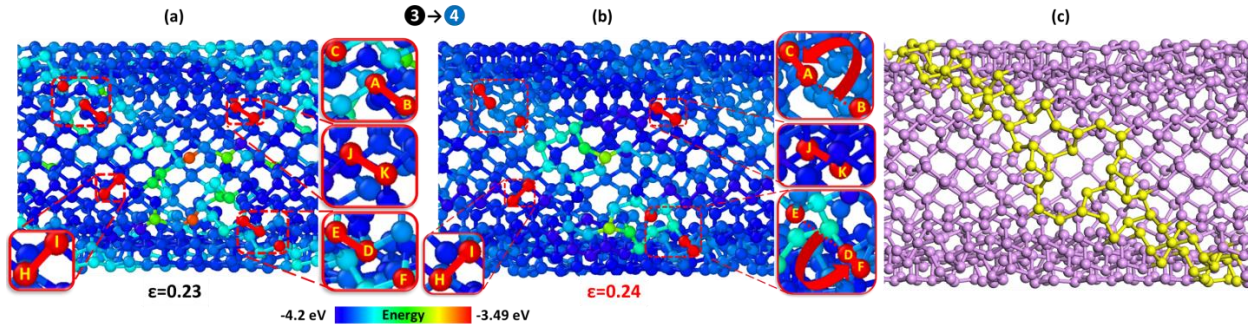

Figure 7: Two successive atomistic configurations taken before (a) and after (b) the second drop (❸→❹) in the strain energy at ε=0.23 and ε=0.24, respectively. Color indicates energy per atom. Energy range is specified by color bar at the bottom. The highlighted triplets of atoms {A, B, C} and {D, E, F} are involved in bond breaking-and-remaking process, while the bonds connecting the {H, I} and {J, K} atoms remain intact. The highlighted atoms are shown in the zoomed-up inset plots. (c) Atoms with the modified bond connectivity located around the divacancy at ε=0.24 (and the bonds connecting them) are highlighted.

Although the energy of atoms with modified bond connectivity is somewhat higher than those with original bond connectivity (see Figure 7(a)), the overall effect of the bond formation and bond relaxation leads to a substantial drop in the strain energy. To show that the strain energy decreases after the second transition, we randomly selected a few outer shell atoms forming the new bonds (see the highlighted atoms A, B, C, D, E and F in the zoomed-up inset plots of Figure 7(a, b)). Before the transition, the A-B and D-E atoms are connected by bonds with the length of $l_{|AB|}$=2.61Å and $l_{|DE|}$=2.59Å, respectively. These bonds are relatively strained since their length exceeds the equilibrium length ($l_b$=2.51Å [16]) of the outer shell bonds at zero strain in a pristine AC PNT with the same diameter. After the transition, the new inter-pucker bonds |AC| and |DF| are formed; their length is $l_{|AC|}$=2.47Å and $l_{|DF|}$ =2.49Å, respectively. The length of the new bond is closer to the equilibrium length; therefore the strain energy associated with the new bonds is reduced. Similarly, the length of the bonds retaining the original bond connectivity is partially adjusted. The two arbitrarily selected outer shell bonds connecting H-I and J-K atoms are highlighted in Figure 7(a, b). The bond length of |HI| and |JK| bonds before and after the second transition is $l_{|HI|}$: 2.57Å→2.53Å and $l_{|JK|}$: 2.64Å→2.57Å, respectively. The length of the selected bonds is closer to the equilibrium one; therefore, the strain energy associated with the bonds is also reduced. The sharp drop in the strain energy at the second transition is a cumulative effect produced by the atoms, which locally change their bond connectivity, and by the bonds, which partially relax after the structural transition.

An additional increase in the applied tensile strain ultimately leads to the fracture of the defective AC PNT at a failure strain $\varepsilon_f$=0.32 (see the final drop in the strain energy at the ❺→❻ transition in Figure 6(a)). The most strained bonds in the AC PNT are located around the divacancy (see the bonds connecting the red atoms in Figure 8(a)). The length of these bonds increases rapidly under applied tensile strain. Near the failure strain, the most strained bonds are on the brink of rupture: in particular, the length of |AC| bond is approaching the bond length limit (see Figure 6(b)). At the failure strain |AC| bond breaks first, triggering the subsequent rupture of the neighboring bonds (see Figure 8(b)). The outer and inner shell bonds are broken (see the highlighted bonds in Figure 8Figure 9(c)), while the bridge bonds remain intact. As a result, a crack with armchair edges is formed (see Figure 8(d)). The crack starts at the divacancy site, passes through the adjacent rings, and goes along the tube circumference perpendicularly to the direction of the applied tensile strain (see Figure 8(d)). Eventually, the defective AC PNT fractures into two halves abruptly, releasing the remaining strain energy.

The critical strain of the defective AC PNTs is considerably reduced in comparison to $\varepsilon_{cr}$ of the pristine PNTs (see Figure 8(f)). Monovacancies and divacancies have a rather similar effect on the critical strain. The largest reduction in the critical strain from $\varepsilon_{cr}\approx0.3$ to $\varepsilon_{cr}\approx0.17$ (almost by $\Delta\varepsilon/\varepsilon_0\approx$-50%) is due to DV(585)A divacancies. The effect of monovacancies on the critical strain is somewhat weaker. The critical strain of both the pristine and defective AC PNTs gradually decreases with the tube diameter (see Figure 8(f)). Since the distance between adjacent puckers is shorter for AC PNTs with the smaller diameters, these nanotubes are partially reinforced by stronger interactions of neighboring puckers. Hence, the smaller the tube diameter is, the larger the critical tensile strain is (see Figure 8(f)).

The critical stress as a function of the tube diameter is shown in Figure 8(e) for the pristine and defective AC PNTs. The critical stress of the pristine AC PNTs is within the range from $\sigma_{cr}\approx20.5$GPa to $\sigma_{cr}\approx21.5$GPa. The introduction of vacancies reduces the critical stress to the range from $\sigma_{cr}\approx16$GPa to $\sigma_{cr}\approx19$GPa. The effect of the hosted vacancies on the critical stress (reduction by $\Delta\sigma/\sigma_0\approx$-25%) is less significant than on the critical strain (reduction almost by $\Delta\varepsilon/\varepsilon_0\approx$-50%). In particular, the critical stress is most noticeably reduced by DV(585)A divacancies, while the effects of DV(585)B divacancies and monovacancies are relatively weaker. The critical stress of the pristine and defective AC PNTs decreases with the tube diameter (see Figure 8(e)).

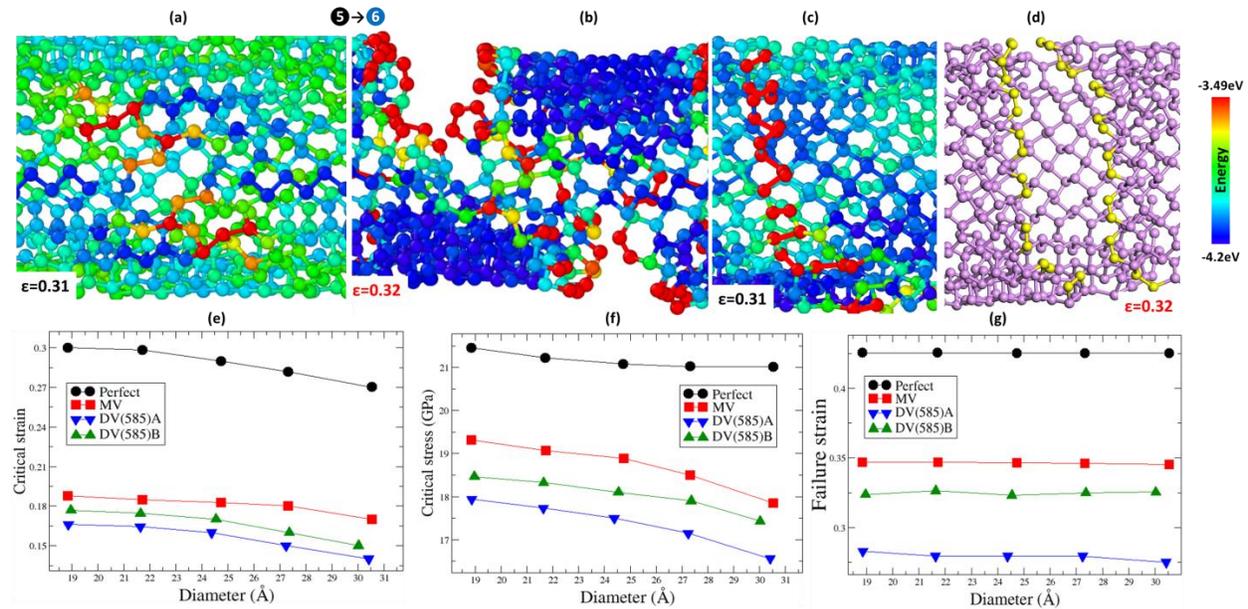

Figure 8: (a, b) Two successive atomistic configurations in an AC PNT (D = 24.5 Å) with a DV(585)B divacancy taken before (a) and after (b) the last drop (❺→❻) in the strain energy at ε=0.31 and $\varepsilon_f$=0.32, respectively. (c) The crack pathway: the bonds rupturing simultaneously at $\varepsilon_f$ are highlighted by red. In (a-c), color indicates energy per atom, while energy range is specified by color bar on the right. (d) The atomistic configuration around the crack: the atoms with broken bonds forming arm-chair edges are highlighted. (e-g) The critical strain (e), critical stress (f) and failure strain (g) for the pristine (black circles) and defective AC PNTs with monovacancies (red squares), DV(585)A (blue triangles down) and DV(585)B (green triangles up) divacancies as a function of tube diameter.

The failure strain as a function of tube diameter is shown in Figure 8(g). The failure strain is higher than the critical one, since after the critical strain the AC PNTs undergo a series of the structural transformations until they finally fracture at a significantly higher tensile strain. The failure strain for both the pristine and defective AC PNTs is almost insensitive to the tube diameter (see Figure 8(g)).

**Deformation and failure of ZZ PNTs**

Finally, we studied the tensile deformation and failure of defective ZZ PNTs subjected to a uniform uniaxial tensile strain. The main stages of the tensile deformation of defective ZZ PNTs are analogous to those of pristine ones [16]: the initial bond stretching is accompanied by the partial bond rotation (see Figure 9(a, b)), which is terminated by bond rupture initiated in the vicinity of a hosted vacancy (see Figure 9(c)). Here, we only present the results for the deformation and failure of ZZ PNTs with a DV(585)A divacancy since the results obtained for ZZ PNTs with a monovacancy and a DV(585)B divacancy are reasonably similar.

To clarify the fracture mechanism, we selected a group of atoms located around the DV(585)A divacancy (see highlighted A, B, C and D atoms in Figure 9(d)) and monitored the bond length of the bonds connecting them as a function of applied tensile strain. As can be seen in Figure 9(e), the length of the bridge bond |BC| (which is oriented along the direction of the applied strain) increases especially fast in a vicinity of a critical strain, in comparison to the length of the two shell bonds |AB| and |CD| (which are oriented at an angle with the direction of the applied strain). The |BC| bond ruptures at the critical strain ($\varepsilon_{cr}$=0.34) causing failure of the neighboring bonds (see Figure 9(c)). Bond breaking, initiated at the failure strain in the vicinity of the divacancy, propagates along the tube circumference and eventually ends in complete brittle-like failure of the ZZ PNT.

In order to examine the effect of the tube diameter on the failure mechanism, we measured the length of the |BC| bond as a function of applied tensile strain in ZZ PNTs with different diameters. As can be seen in Figure 9(f), the larger the tube diameter is, the more stretched the |BC| bond length near the failure strain is. The stress-strain curves for the defective ZZ PNTs with the various diameters are shown in Figure 9(h).

The failure strain of the pristine and defective ZZ PNTs as a function of tube diameter is shown in Figure 9(i). Vacancies significantly reduce the failure strain of the ZZ PNTs. Thus an introduction of a monovacancy reduces the failure strain of the ZZ PNT with the largest diameter (D=24.3Å) from $\varepsilon_{f0}$=0.62 (pristine nanotube) to $\varepsilon_f$=0.34 (defective nanotube) almost by $\Delta\varepsilon_f/\varepsilon_{f0}$≈-50%. The two types of divacancies have a slightly weaker effect on the failure strain. The failure strain of the ZZ PNTs slightly decreases with the tube diameter (see Figure 9(i)).

The vacancies also reduce the failure stress of ZZ PNTs (see Figure 9(g)), although their impact is less dramatic than that on the failure strain. As an example, a monovacancy reduces the failure strain of the ZZ PNTs with the largest tube diameter from $\sigma_{f0}$=6.6GPa (pristine nanotube) to $\sigma_f$=5.9GPa (defective nanotube) only by $\Delta\sigma_f/\sigma_{f0}$≈-10%. The effect of the divacancies on the failure stress is comparable to that of the monovacancy. The failure stress for the pristine and defective ZZ PNTs slightly increases with the tube diameter (see Figure 9(g)).

The weak, but noticeable effect of the tube diameter on the failure strain and failure stress of the defective ZZ PNTs is due to initial pre-straining of the bridge bonds [16], especially of the |BC| bonds around the divacancy (see Figure 9(d)). The initial length of the |BC| bond (at zero strain) increases with the tube diameter (see Figure 9(f)): the larger the tube diameter is, the more pre-strained the bond is. Due to initial pre-straining, the |BC| bond length in the wider nanotube increases faster with applied tensile strain (see Figure 9(f)). Subsequently, the |BC| bond of the ZZ PNT with the largest tube diameter reaches first a specific bond length ($l_s$≈2.6Å) at which the bond elongation switches from a linear to a non-linear regime (see Figure 9(f)). Similarly, the |BC| bond reaches first the bond length limit ($l_{max} \approx 3$Å) at which it breaks. Hence, the failure strain decreases with the increasing tube diameter, although the effect is minor. The failure stress increases with the tube diameter, since the |BC| bonds in the ZZ PNTs with larger tube diameters enter the non-linear regime at a lower tensile strain. Given that a higher tensile stress is required to strain the bonds in the non-linear regime, the bond failure occurs accordingly at a higher failure stress. The increase in the failure stress with the tube diameter is also relatively moderate.

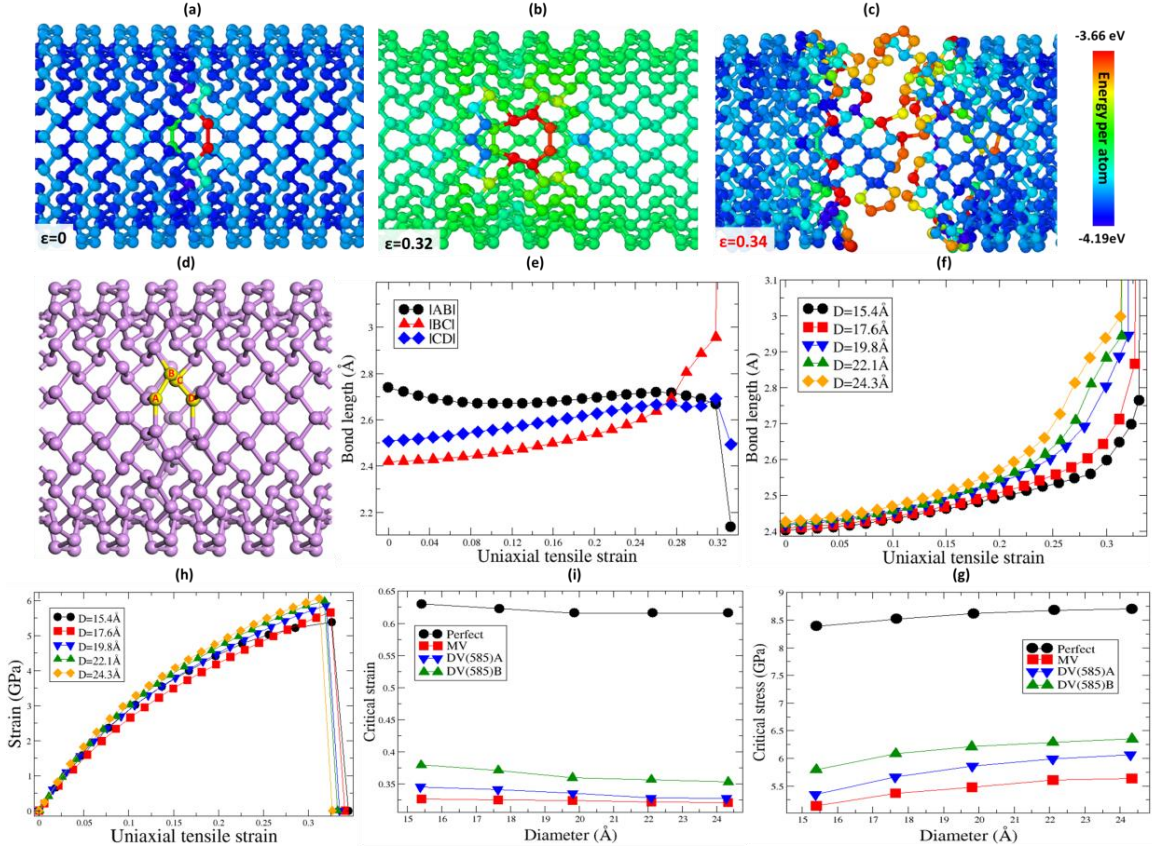

Figure 9: (a-c) The deformation and failure of ZZ PNTs with a DV(585)A type divacancy under applied tensile strain. The images are taken at ε=0 (a), ε=0.32 (b) and $\varepsilon_f$=0.34 (c). Color specifies energy per atom. Energy range is indicated by the color bar on the right. The tube diameter is D=22.1 Å (d) The atomistic configuration around a DV(585)A divacancy: the selected A, B, C and D atoms are highlighted. (e) The length of the bonds connecting A-B (black circles), B-C (red triangles), and C-D atoms (blue diamonds) as a function of applied tensile strain (f) The length of the |BC| bond as a function of the applied tensile strain for ZZ PNTs with different diameters. (h) Stress-strain curves of defective ZZ PNTs with a DV(585)A divacancy. The tube diameter is varied from D=15.4 Å to D=24.3 Å (i, g) The failure strain (i) and failure stress (g) as a function of tube diameter for the pristine (black circles) and defective ZZ PNTs with a monovacancy (red squares), a DV(585)A (blue triangles down) and a DV(585)B (green triangles up) divacancy.

## 4. Conclusions

Using the DFTB method, we investigated the mechanical properties, deformation and failure of PNTs with monovacancies and two variants of divacancies subjected to uniaxial tensile strain. The atomistic configurations around vacancies were examined and the vacancy formation energies were calculated as a function of the tube diameter. In AC PNTs, a DV(585)B divacancy possesses the lowest vacancy formation energy, while a DV(585)A divacancy possesses the highest one. In ZZ PNTs, the vacancy formation energy of a monovacancy is the lowest one, while that of a DV(585)A divacancy is the highest one. In AC PNTs, the vacancy formation energy decreases with the increasing tube diameter, in contrast in ZZ PNTs, it increases with the increasing tube diameter. For AC PNTs with large diameters, the vacancy formation energies of DV(585)B divacancy and monovacancy become comparable.

We also studied the effect of vacancies on the Young's modulus, Poisson's ratios and flexural rigidity of defective PNTs. The Young's modulus is moderately reduced by vacancies: in AC PNTs, the effect of a DV(585)A divacancy is the strongest one ($\Delta Y/Y_0 \approx -8\%$), while in ZZ PNTs, a monovacancy reduces the elastic modulus most ($\Delta Y/Y_0 \approx -7\%$). The Young's modulus decreases with the tube diameter in AC PNTs, but increases in ZZ PNTs. Both the radial and thickness Poisson's ratios increase when vacancies are hosted in PNTs. The flexural rigidity moderately increases in defective PNTs, in particular by divacancies.

We also investigated the tensile deformation and failure mechanism of defective PNTs. Under applied tensile strain, AC PNTs undergo a sequence of structural transformations, leading to the ultimate failure. Due to the transverse contraction, the distance between adjacent puckers decreases with applied tensile strain. At a critical strain, a number of the intra-pucker bonds rupture and new inter-pucker bonds form over the entire tube shells in pristine AC PNTs. In defective AC PNTs, however, the most strained bonds rupture around a vacancy and form new inter-pucker bonds at a lower critical strain. As a result, a sequence of the intra-pucker bond-breaking and inter-pucker bond-remaking events is initiated in the vicinity of the vacancy. The new inter-pucker bonds in defective AC PNT are located within a narrow region surrounding the vacancy. Upon further stretching, the tensile deformation of AC PNTs is interrupted again by a subsequent formation of the new inter-pucker bonds. This sequence of the structural transformations terminates at the final failure of defective AC PNTs. The failure begins with the rupture of the overstretched bonds around the vacancy at the failure strain (which is almost twice as large as the critical strain). The critical strain of defective AC PNTs is markedly reduced by vacancies: the largest reduction due to a divacancy ($\Delta\varepsilon/\varepsilon_0 \approx -50\%$). The effect of vacancies on the critical stress is relatively weaker: the largest reduction is also due to a divacancy ($\Delta\sigma/\sigma_0 \approx -25\%$). Both the failure strain and failure stress slightly decrease with the tube diameter.

In contrast to AC PNTs, the failure of ZZ-PNTs occurs in a brittle-like manner when the most strained bonds around a vacancy rupture, forming a crack running along the tube circumference. Vacancies significantly reduce the failure strain of ZZ PNTs, with the largest reduction coming from a monovacancy ($\Delta\varepsilon/\varepsilon_0 \approx -50\%$). The failure stress of ZZ PNTs is also reduced by vacancies, although the effect is less dramatic compared to the failure strain: the largest reduction ($\Delta\sigma/\sigma_0 \approx -10\%$) is also due to a monovacancy, while the effect of divacancies is comparable. The failure strain of defective ZZ PNTs slightly decreases, whereas the failure stress to some extent increases with the tube diameter.

It is interesting to compare the failure behavior of defective PNTs with that of defective graphene [42], its allotropes [43,44] and carbon nanotubes [45,46]. Similar to defective graphene [42] and CNTs [45,46] the first sign of fracture in defective PNTs is the rupture of the overstretched bonds around vacancies. All the broken P-P and C-C bonds are oriented parallel to the direction of applied tensile strain. Like in defective graphene [42] and CNTs [45,46], cracks (oriented normally to the loading direction) are generated around the broken bonds: once initiated the cracks continue to grow until the ultimate failure of defective PNTs. The fracture of defective ZZ PNTs is brittle-like similar to the fracture of defective graphene [42] and CNTs [45,46]. In contrast, the fracture of defective AC PNTs, which proceed through a series of local structural transformations, is ductile-like. Ductile fracture is typical for graphene allotropes (undergoing an order-disorder transition under tensile strain at low temperatures [43]), and defective graphene and CNTs at elevated temperatures [42,45–47].

PNTs can be potentially used as building blocks in nanoelectromechanical systems. Since point defects in PNTs in general cannot be avoided, thus it is important to understand the mechanical properties and the deformation and failure mechanisms of defective PNTs. Such understandings enable one to impose specific restrictions on the mechanical strains and stresses in nanoelectromechanical systems containing defective PNTs. It is expected that the present findings will provide the useful guidelines for the design and fabrication of PNT building blocks and promote their applications in nanoelectromechanical systems.

# 5. Acknowledgements

This work was supported by the A*STAR Computational Resource Centre through the use of its high performance computing facilities, and by a grant from the Science and Engineering Research Council (152-70-00017).

# Citations